\documentstyle[psfig]{lamuphys}
\makeatletter
\let\chapter\hid@chapter
\makeatother
\begin{document}
\pagenumbering{arabic}
\title{The Physical Origin of Galaxy Scaling Relations}

\author{Simon\,D.M.\,White}

\institute{Max-Planck-Institut f\"ur Astrophysik,
Karl-Schwarzschild-Stra{\ss}e 1,\\ D-85740 Garching bei M\"unchen, Germany}

\maketitle

\begin{abstract}
A standard paradigm is now available for the recent evolution ($z < 10$)
of structure on galactic and larger scales. Most of the matter is
assumed to be dark and dissipationless and to cluster
hierarchically from gaussian initial conditions. Gas moves under
the gravitational influence of this dark matter, settling
dissipatively at the centres of dark halos to form galaxies. The
evolution of the dark matter component has been studied extensively
by N-body simulations. The abundances, density profiles, shape
and angular momentum distributions and the formation histories of the 
dark halo population can all be predicted reliably for any hierarchical
cosmogony. The systematic variation of these properties with halo
mass can produce scaling relations in the galaxy population.
Simple hypotheses for how galaxies condense within dark halos
lead to characteristic luminosities, sizes, and spins which are close
to those of real spiral and elliptical galaxies. Furthermore, correlations
similar to the Tully-Fisher, Faber-Jackson and luminosity-metallicity
relations arise quite naturally. A quantitative explanation of the 
fundamental plane of elliptical galaxies appears within reach.
\end{abstract}
\section{Introduction}

The current popularity of the Cold Dark Matter (CDM) model and its 
variants stems from a variety of sources. Most of the mass in the
Universe appears to be in a dark collisionless form 
which is concentrated towards galaxies but extends far beyond their 
visible boundaries. If this matter has interacted only gravitationally
since early times it is possible to reconcile the very small observed 
amplitude of fluctuations in the Microwave Background with the existence
of massive nonlinear structures in the present Universe. Furthermore,
the large-scale distribution of galaxies looks quite similar to
the patterns which result from the gravitational amplification
of gaussian density fluctuations. This is a simple and natural initial
condition in CDM-like models, and in such models the galaxies are 
indeed expected to trace the dark matter distribution on large scales.

The idea that hierarchical clustering under gravity has given rise
to galactic and larger structures predates the CDM model
(e.g. Peebles 1980), and almost two decades of simulation using
N-body methods have provided a reasonably complete understanding of the
structure produced by this process. If galaxy formation results from
the dissipative collapse of gas within the potential wells provided
by dark halos, then it is the internal structure of such halos
and their formation history which must regulate the global properties
of galaxies. In recent years there has been substantial progress in
understanding the predictions of hierarchical clustering models
for these and other aspects of halo structure. In the next section I
summarise the aspects of this work most relevant for a discussion
of galaxy scaling relations.

The structure and evolution of dark halos may determine the mass and
angular momentum of the material available for galaxy formation, as
well as the rate of interactions between galaxies.
The global properties of galaxies must in addition depend
on how gas cools to form dense clouds, on how
star-formation proceeds in such clouds, and on how this star formation
affects the surrounding material through the injection of heavy elements and
energy. These processes interact in a highly nonlinear way and involve
a very wide range of scales; there is little hope of simulating them 
realistically. If, however, they are parametrised by an appropriate set
of efficiencies, to be assigned physically reasonable values
by comparison with available observational or simulation data, then it
is possible to make simple ``semi-analytic'' models which can
predict a very wide range of properties of the galaxy
population for any specific cosmogony, for example any of the popular
CDM variants (e.g. luminosity functions, colours, sizes, morphologies, gas
contents, and the dependence of all of these on environment and
redshift). In sections 3 and 4 I discuss how such models can be
used to predict galaxy scaling relations, and I emphasise, in
particular, the inferences which can be drawn from the observed
tightness of some of these relations.

\section{Structure and Evolution of Dark Halos}

N-body simulations have provided a good understanding of the 
structure and evolution of the dark halos which form through
hierarchical clustering of dissipationless matter. For example,
the angular momentum of dark halos, best characterised by the
dimensionless spin parameter,
\begin{eqnarray}
\lambda\equiv J|E|^{1/2}/GM^{5/2},
\end{eqnarray}
where $J$, $E$ and $M$ are the total angular momentum, energy and mass
of the halo, is found to have a broad distribution with a median near 
$\lambda\sim 0.04$. This distribution appears ``universal'' in the 
sense that it has no strong dependence on the mass of the halo or on 
the parameters of the particular cosmogony in which clustering occurs 
(e.g. initial power spectrum, $P(k)$, cosmic density, $\Omega$, 
cosmological constant, $\Lambda$,...; see Barnes \& Efstathiou
1987, Frenk et al 1988, Warren et al 1992, Cole \& Lacey 1996).
For a cold rotationally-supported self-gravitating disk one
finds $\lambda\sim 0.4$. Hence $\lambda\sim 0.04$ implies a system
in which rotation velocities are an order of magnitude smaller
than needed for centrifugal support. Any gas component in such a system must
shrink in radius by a similar factor if it is to make a centrifugally 
supported disk (see below). Within individual dark halos rotational
streaming usually varies quite weakly with radius, but there are large
variations from halo to halo.

The axial ratios of dark halos have also been extensively studied
and also appear to show a broad distribution which depends at most
weakly on mass or cosmogony (Frenk et al 1988; Warren et al 1992;  
Cole \& Lacey 1996). Nearly spherical halos are quite uncommon.
There is a slight preference for near-prolate over near-oblate shapes,
and major-to-minor axis ratios in excess of two are common. It is
interesting to ask whether such a distribution is consistent with the
fact that deviations from axisymmetry in observed disks are typically 
quite small (e.g. Rix \& Zaritsky 1995). I will not pursue this
question further here.

A third regularity in the structure of dark halos has emerged only
recently. Navarro et al (1996, 1997; hereafter NFW) studied halo 
density profiles in a wide variety of hierarchical cosmogonies.
Their work is distinguished from earlier studies in that they
simulated the evolution of each halo separately. This allowed them
to set the resolution limits in mass and in linear scale to be 
constant fractions of the  
characteristic mass and radius of each halo, even though these
characteristic values ranged over several orders of magnitude. NFW
found the remarkable result that the spherically averaged density
profiles of halos of {\it all} masses in {\it all} the cosmogonies
they considered could be adequately represented by a suitable scaling
of the same analytic form:
\begin{eqnarray}
\rho(r)/\rho_{crit} = \delta_cr_s^3/r(r+r_s)^2.
\end{eqnarray}
In this formula $r_s$ sets the ``core'' radius of the halo and
$\delta_c$  is its characteristic density in units of the critical 
density, $\rho_{crit}$. Thus the inner regions have a
density cusp with $\rho\propto 1/r$ while at larger radii the profile
steepens towards $\rho\propto 1/r^3$. The bounding radius of a 
virialised halo is
conventionally defined as the radius $r_{200}$ within which the mean
density is 200 times the critical value; the ``total'' halo mass is
then the mass within this radius $M_t$. Defining the
concentration of a halo to be $c\equiv r_{200}/r_s$, one immediately
finds $c$ to be determined implicitly from
\begin{eqnarray}
\delta_c = {200\over 3}{c^3\over [\ln(1+c)-c/(1+c)]},
\end{eqnarray}
and, of course, we have
\begin{eqnarray}
M_t={4\pi\over 3}~200\rho_{crit}~(cr_s)^3.
\end{eqnarray}
NFW found that for any particular hierarchical cosmogony the two
parameters $\delta_c$ and $r_s$ are strongly correlated with each
other and so with $M_t$. This correlation is always in the sense that
lower mass halos have higher characteristic densities and so greater
concentrations. It turns out that this correlation can be understood
as a reflection of the fact that smaller mass halos form 
earlier, and indeed, for a suitable definition of the formation epoch
of a halo $z_f$, NFW showed that all the halos in all their cosmogonies
obey the simple relation
\begin{eqnarray}
\delta_c \approx 5\times 10^3 \Omega_0(1+z_f)^3.
\end{eqnarray}
To a good approximation it seems that equilibrium dark halos in all
hierarchical cosmogonies have similar density profiles and furthermore
that the characteristic density of a halo is just proportional to the
density of the universe at the time it formed. It is hard to imagine a
simpler situation.

Of course, the properties of the galaxies within a dark halo depend
not only on its current structure but also on the details of its
formation history. There has been substantial progress over the last
five years in understanding how individual nonlinear objects are built
up by hierarchical clustering. This is primarily a result of the
discovery that extensions of the original argument of Press \&
Schechter (1974) can provide a remarkably detailed and accurate
description of the statistics of merging and accretion in N-body
simulations of hierarchical clustering (Bond et al 1991; Bower 1991; 
Lacey \& Cole 1993,1994). Indeed, these formulae provide a basis for 
Monte Carlo realisations of the full merging tree which
describes how any particular object, for example a rich cluster, is
built up by successive merging of smaller systems (Kauffmann \& White
1993). 

Armed with such a tree one can attempt to model all the
additional processes which determine the
galaxy population within a dark halo (gas cooling, star formation,
energy injection from young stars, chemical enrichment, stellar
population evolution, galaxy (as opposed to halo) merging, etc.). 
A major success of recent galaxy formation studies has been
the demonstration that even very simple physical models for
these processes lead to explanations not only for the luminosities, 
colours, morphologies, metallicities and abundances of galaxies, and
for scaling relations between these properties, but
also for the fact that $10^{15}$M$_\odot$ halos typically 
contain many bright early-type galaxies while $10^{12}$M$_\odot$ 
halos typically contain a single central spiral and a few 
satellites (Kauffmann et al 1993; Baugh et al 1996a). Furthermore,
since such models automatically specify the full history of the galaxy
population, they can be compared directly with observational
indicators of galaxy evolution, for example with counts and redshift
distributions of faint galaxies (Cole et al 1994; Kauffmann et al 1994; Heyl
et al 1995; Baugh et al 1996b) or with the properties of damped Ly$\alpha$
systems in QSO spectra (Kauffmann 1996a). The results so far are
encouraging, and it seems that a reasonably complete, if schematic,
picture of galaxy formation is now in place. In the next two sections
I discuss the scaling relations this picture predicts for
disk and elliptical galaxies.

\section{Scaling Relations for Disks}

The defining characteristic of galaxy disks is that they are made of
stars which are almost all on near-circular orbits confined close to
the disk plane. Since it appears impossible to create a 
thin centrifugally supported disk without very substantial
dissipation, one draws two immediate conclusions:\hfil\break
(i) Galaxy disks were assembled while still gaseous -- their stars 
were all formed {\it in situ}. Of course, this does not preclude 
disk growth through gas infall after the formation of many of
the stars. \hfil\break
(ii) galaxy disks cannot have been violently disturbed since formation
of the bulk of their stars, otherwise they would no
longer be thin.\hfil\break
Another critical observation is that the outer rotation curves of most
spirals are approximately flat and appear to supported primarily by
the gravity of their dark halos. This suggests that the properties of 
disks may be determined by those of their dark halos.

The standard model for disk formation was set out by Fall \&
Efstathiou (1980) in an extension of the ideas of White \& Rees 
(1978). After a dark halo comes to equilibrium, much of its baryonic 
material is supposed to remain as diffuse gas with a distribution 
similar to that of the dark matter. Subsequently this gas radiates its
binding energy but retains its angular momentum and so flows inwards 
until it settles into a rotationally supported disk. Fall \&
Efstathiou showed that an extended dark halo is required, and that 
little angular momentum can be lost if disks similar to observed 
spirals are to form. Their scheme is the basis of most recent 
modelling of spiral galaxy formation (e.g. Kauffmann 1996a;  Dalcanton
et al 1997) but has not yet been shown to work in any numerical 
simulation of hierarchical galaxy formation. The difficulty is that 
inclusion of feedback from young stars is critical. Without it gas 
cools into small dense clumps at early times, and these lose most 
of their angular momentum as they merge at the centres of massive 
dark halos; the resulting disks are then too small to represent real
galaxies (Navarro \& Benz 1991; Navarro et al 1995; Navarro \& 
Steinmetz 1997). 

Let me work through a simple example to show how this scheme can be
used to derive scaling relations for galaxy disks. If we model a halo
as a singular isothermal sphere of circular velocity $V_c$, then its
mass, kinetic energy and angular momentum within $r_{200}=V_c/10H(z)$ 
can be written as
\begin{eqnarray}
M_t= {V_c^3\over 10GH(z)},~~~E=M_tV_c^2/2,~~{\rm and}~~J_h=\sqrt{2}
\lambda M_tr_{200}V_c,
\end{eqnarray}
where $H(z)$ is the Hubble constant at the redshift when the halo
is identified and its central disk is made. Assume a fraction $F$ of 
the mass of the halo is in the
form of gas with the {\it same} specific angular momentum as the dark
matter. Assume further that this gas sinks to the centre conserving 
its angular momentum and forms an exponential disk of mass $M_d$, 
central surface density $S_0$ and scale radius $r_d$. If we neglect 
the contribution of the self-gravity of the disk to its rotation curve
then we find
\begin{eqnarray}
M_d={FV_c^3\over 10GH(z)},~~~r_d={\lambda V_c\over
10\sqrt{2}H(z)}~~{\rm and}~~S_0={10FV_cH(z)\over\pi\lambda^2G}.
\end{eqnarray}
These relations have some immediate consequences. If the stellar
mass-to-light ratio of disks is assumed to be a constant value
$\Upsilon(z)$ at each redshift, then the first relation gives
a Tully-Fisher-like relation, $L\propto V_c^3$, which is {\it
independent} of $\lambda$. On the other hand $r_d$ and $S_0$
depend strongly on $\lambda$; slowly rotating halos produce compact 
and high surface brightness disks. This is encouraging because the
exponent of the observed Tully-Fisher relation is not far from 3, and 
furthermore this relation appears to hold independent of galaxy
surface brightness (de Blok \& McGaugh 1996; Tully, this meeting). In
addition, the proportionality constant seems reasonable. If, following
McGaugh \& de Blok (1997), we adopt $\Upsilon_B=2.5h$, then the zero-point
of the observed T-F relation, $L_B=1.5\times 10^{10}h^{-2}L_\odot$
at $V_c=200$km/s (e.g. Strauss \& Willick 1995), agrees with the
prediction provided $F=0.02H_0/H(z)$, i.e. $F\approx 0.02$ if disks are
assembled near $z=0$ and $F\approx 0.05$ if disks are assembled near
$z=1$.

The predicted characteristic sizes of disks also seem reasonable.
For a ``typical'' halo with $\lambda=0.05$ and $V_c=200$km/s the
predicted scale radius is $r_d=7H_0/H(z)~h^{-1}$kpc, or $R_d\approx
7h^{-1}$kpc for assembly near $z=0$ and $R_d\approx 3h^{-1}$kpc for
assembly near $z=1$. Notice that the redshift dependence in these
equations is quite strong. It does not appear possible to make
substantial numbers of big disks at high redshifts. Thus if
damped Ly$\alpha$ absorbers in QSO spectra at $z\sim 3$ are indeed
equilibrium disk systems with circular velocities of order 200km/s,
then they must be quite small, $r_d\sim 1$ to 2 kpc, if they are to
be explained in a hierarchical clustering model. Notice also, as
mentioned above, that there cannot be much transfer of angular
momentum from gas to dark matter during disk formation, otherwise the
resulting disks will be too small for {\it any} assumed redshift of 
assembly.

The strong $\lambda$-dependence of $r_d$ and $S_0$ together with the
broad $\lambda$-distribution resulting from hierarchical clustering
implies that galaxy disks are predicted to have
a wide range of sizes and surface brightnesses at any given luminosity
or circular velocity. A recent discussion of the observational data 
by Dalcanton
et al (1997) suggests that this may indeed be the case. ``Disks'' 
formed from the low $\lambda$ tail of the distribution are predicted 
to be so compact, however, that they should perhaps
be identified with observed spheroids. In these objects the
baryonic component should dominate strongly over the dark
matter, and this may, perhaps, lead
to violent instabilities which prevent thin disk formation.

A final important issue concerns the tightness of the 
observed T-F relation. This obviously implies some considerable
uniformity in the formation of disk galaxies. As we have seen, the
broad spin distribution does not, 
of itself, induce scatter. Variations in assembly 
time can do so through the $H(z)$ dependence of $M_t$. In 
combination with the size constraints already discussed, this suggests
that most disks were assembled well after $z=1.$ Variations
in the actual structure of halos of given mass and assembly 
epoch must also be sufficiently small to avoid excessive scatter
in the $M_t-V_c$ relation. For the halos simulated by Navarro et al
(1996, 1997) this relation is indeed tight enough. Finally, small scatter is
required in the fraction $F$ of the halo mass which condenses into a
disk, in the disk mass-to-light ratio $\Upsilon$, and in the disk 
contribution to the observed $V_c$ values (which will vary with 
$\lambda$). The observed colours of disk galaxies are quite uniform,
suggesting that $\Upsilon$ may not vary too much, and recent observations
favour small $\Upsilon$ values, thus helping to satisfy the last condition
(e.g. McGaugh \& de Blok 1997). Since the required $F$-values are smaller
than observed in galaxy clusters (e.g. White and Fabian 1995), the 
uniformity of $F$ suggests that some feed-back process lowers the 
condensation efficiency in a way which depends only on $V_c$. 
Substantial feed-back appears necessary to account for 
the apparent global inefficiency of galaxy formation (e.g. White \& 
Rees 1978; White \& Frenk 1991) and a variation with $V_c$ can induce
a metallicity-luminosity relation (Larson 1972; Dekel \& Silk 1986). In 
particular, feedback from star formation in CDM-like cosmologies can
plausibly explain the observed metallicities both of present-day disks
and of high redshift damped Ly$\alpha$ systems (Kauffmann 1996a) 

A more careful analysis of many of the ideas in this section,
together with applications to specific hierarchical cosmologies can
be found in Mo et al (1997) and Dalcanton et al (1997). The latter
paper compares its predictions in some detail with the observed
sizes and surface brightnesses of disk galaxies. 

\section{Scaling Relations for Ellipticals}

The properties of elliptical galaxies, particularly those of
ellipticals in rich clusters, show some remarkable regularities. 
Most have very nearly elliptical isophotes and a luminosity profile
which is well described by de Vaucouleurs' empirical fitting
function. There is a tight relation, known as the fundamental plane,
between the characteristic size of a galaxy, its total luminosity,
and its central velocity dispersion. In addition there are tight
relations between the luminosities of ellipticals and their colours
and metallicities. The simplest interpretation requires\hfil\break
(i) that all ellipticals are made of old stars,\hfil\break
(ii) that they all formed in a similar way,\hfil\break
(iii) that the initial mass functions of their stellar populations
(and so their $M/L$ ratios at given age) are similar or at least 
vary only slowly with mass, and\hfil\break
(iv) that their metallicity increases (and so their colour reddens) 
systematically with mass.\hfil\break
The fundamental plane then reflects the virial relation
$M\sim R \sigma^2$ with a slight tilt arising from the systematic
variation of $M/L$ with mass. Recent data on the evolution
of ellipticals support this interpretation in that they are consistent
with the fading in luminosity expected for a passively evolving
population of equilibrium galaxies (see other contributions to this 
volume). An indication that the real picture
may be more complex comes, however, from dynamical analyses which 
suggest that much of the mass within the luminous regions of 
ellipticals may in fact be pregalactic dark matter (e.g. Rix et al 1997).

More than twenty years ago Toomre (1977) remarked that star formation
is observed only in galaxy disks, and further that
the final state of pairs of interacting spirals must be something 
resembling an elliptical galaxy. In view of this he suggested that
{\it all} star formation might occur in disk systems, and that
ellipticals might {\it all} be formed by the merger of stellar disks.
Although remaining controversial, these suggestions have gained much
theoretical and observational support since they were made. Direct
simulations of mergers between systems resembling disk galaxies
have shown that they do indeed evolve into objects with a 
structure very like that of ellipticals (e.g. Barnes
1988). Furthermore, a number of transition cases have been found which
seem to demonstrate empirically that merging spirals end up as
ellipticals (e.g. Schweizer 1990). Finally it is still true that
substantial star formation has been seen only in galaxy disks, or in
starbursts either in the nuclear regions of gas-rich galaxies or in 
interacting disk systems.

The strongest objections to Toomre's proposal have come:\hfil\break
(i) from the tight systematic relations between E-galaxy properties
-- tight correlations seem intuitively surprising if ellipticals are 
produced by the stochastic accumulation of smaller units,\hfil\break
(ii) from the fact that ellipticals are denser and more strongly 
bound than spirals -- their progenitors must then have been more
compact and more tightly bound than present-day disks, and\hfil\break
(iii) from the fact that most disk galaxies have central bulges 
which resemble ellipticals in many of their properties -- how could
mergers produce a central ``elliptical'' without disturbing the
surrounding disk.\hfil\break
Semi-analytic models of hierarchical galaxy formation
generally adopt the hypothesis that all star
formation occurs in quiescent or interacting disks, and can address 
the above objections directly
because they keep track of how and where disks grow and of how they merge
together. It is therefore possible to trace the formation history
of each elliptical galaxy, and to ask how it
depends on luminosity and
environment. The first detailed models of this kind were able to
reproduce the characteristic luminosities and colours of ellipticals,
the distribution of bulge-to-disk ratios of spirals, and the
environmental segregation between ellipticals and spirals (Kauffmann
et al 1993; Baugh et al 1996a). Objects with little or no
disk are predicted to occur primarily in clusters and to have
old stellar populations. They form by the merger
of disks which were assembled well before $z=1$ and so were compact
(equ.~7). Present-day disks form late by accretion of new gas onto small
``ellipticals'' produced by the merging of
earlier generations of disks.

\begin{figure} 
\centerline{
\psfig{figure=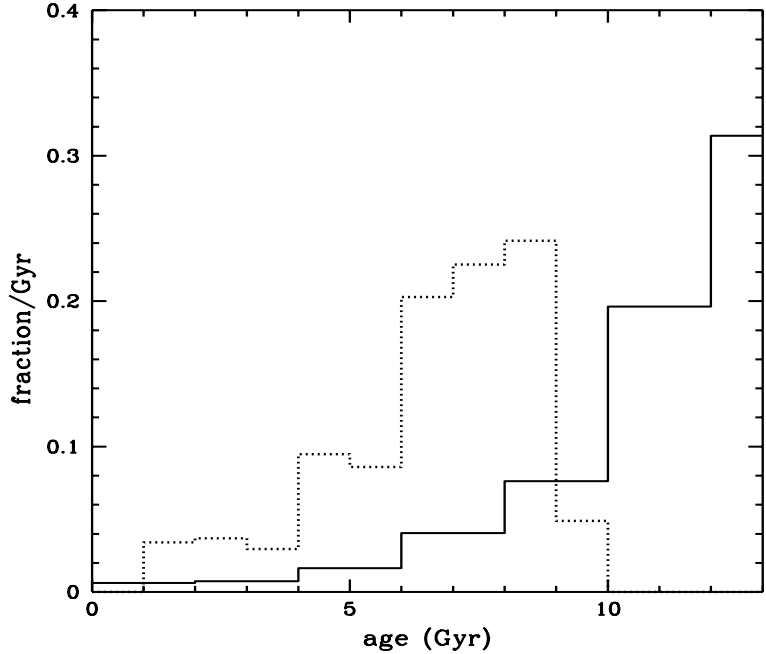,width=10.5cm,height=10.0cm}
}
\caption{The solid histogram gives the distribution of formation times
for the stars in elliptical galaxies in a
$10^{15}$M$_\odot$ cluster. The semi-analytic model assumes a standard
CDM cosmogony normalised to
$\sigma_8=0.67$. The dotted histogram gives the distribution of the
times when these elliptical galaxies underwent their last major merger
(data taken from Kauffmann 1996b). 
}
\end{figure}

In Figure 1 (adapted from Kauffmann 1996b) I illustrate when
star-formation and merging are predicted to occur for cluster
ellipticals in an $\Omega=1$ CDM cosmogony normalised to give
the correct abundance of rich clusters. The modelling scheme assumes
that all objects with disk-to-bulge ratios less than 0.67
are classified as ellipticals, and for this plot the elliptical
population in clusters of mass $10^{15}$M$_\odot$ is analysed. The
solid histogram shows the formation times of the stars which end 
up in these ellipticals. More than 40\% form before $z=3$, about
60\% before $z=2$, and more than 80\% before $z=1$. Very few stars
have formed in these objects over the last few billion years. Thus
cluster ellipticals are predicted to be red and to show
little scatter in their colour-luminosity relation. More detailed
study shows that ellipticals in high-$z$ clusters are predicted
to form their stars earlier on average than present-day
ellipticals, and as a result the scatter in the luminosity-colour relation
remains small out to redshifts of order unity (Kauffmann 1996b).
The dashed histogram in Fig.~1 shows when these ellipticals underwent
their last major merger. This is predicted to be quite late -- more 
than 70\% were assembled after $z=1$. Somewhat later star-formation
and merger times are predicted for ellipticals in groups rather than
clusters, and similar patterns are predicted in other cosmogonies
-- formation is somewhat earlier in low density universes and somewhat
later in $\Omega=1$ cosmogonies with less small-scale power than CDM
(e.g. mixed dark matter).

A natural prediction of hierarchical cosmogonies is that small things
form first. As Figure 2 demonstrates, however, this effect is barely
detectable for ellipticals in clusters. In this plot the mean stellar 
age of ellipticals is shown as a function of their total stellar mass
for the same cosmogony analysed in Fig.~1. Ellipticals of all masses 
are made of old stars, and the decrease in age with increasing mass is
less than the (small) age scatter between galaxies. If metallicity effects
are ignored, the colours of ellipticals are predicted to be essentially
independent of luminosity and to have small scatter (e.g. Kauffmann 
1996b; Baugh et al 1996a). The inclusion of chemical enrichment effects
can plausibly produce the observed colour-luminosity relation
because: (a) more massive ellipticals are predicted to form
from the merging of more massive disks, and (b) as a result of feed-back
effects, the metallicity of disks is predicted to increase strongly
with their mass (e.g. Kauffmann 1996a). The first effect is
illustrated in Fig.~2 which gives the ratio of mean progenitor mass to
final mass as a function of final mass. The mean progenitor mass is 
defined by tagging each star with the mass of the disk galaxy
in which it formed, and then averaging this mass over all the stars in
the final elliptical. The stars in a $2.5\times 10^{12}$M$_\odot$ elliptical
typically formed in disks which were more than ten times as
massive as those which merged to make a $10^{10}$M$_\odot$
elliptical. It will be possible to check whether the resulting
metallicity-luminosity relation reproduces the observed
colour-luminosity plots as soon
as reliable population synthesis models are available for 
a wide range of metallicities.

\begin{figure}
\centerline{
\psfig{figure=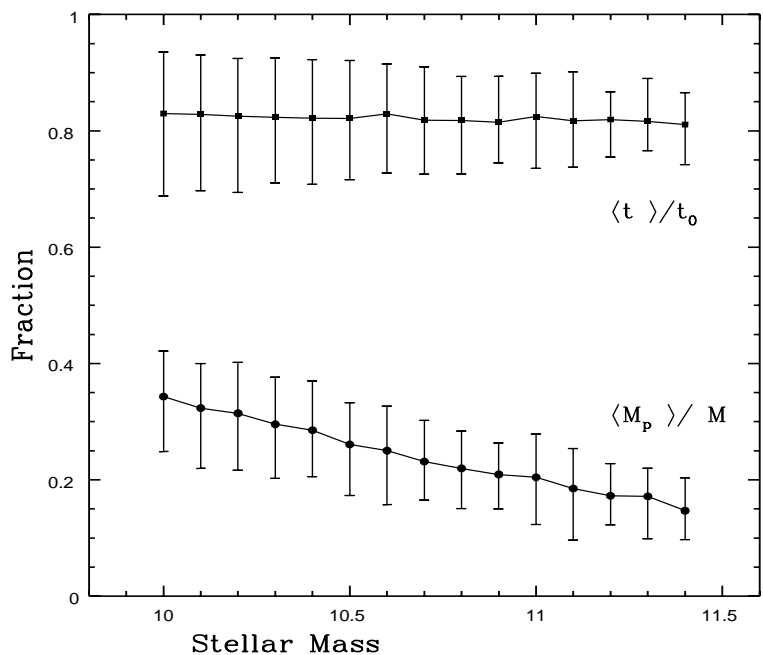,width=10.5cm,height=10.0cm}
}
\caption{The upper points give the mean ages of cluster ellipticals 
(in units of the age of the Universe) as a function of their stellar
mass for the same model plotted in Fig.~1. The lower points give the
mean progenitor masses of these same ellipticals in units of their
total mass. In both cases the error bars join the upper and lower 5\%
points of the galaxy to galaxy scatter in these quantities.
}
\end{figure}

According to these models the stellar population of present-day ellipticals 
formed in compact
disks with scale radii $\sim 1$kpc and circular velocities $\sim 
200$km/s. These disks were among the most rapidly
star-forming objects at $z\sim 2.5$, and should presumably be
identified with the galaxies recently discovered by
Steidel and collaborators (Steidel et al 1996; Giavalisco et al 1996).
The observed objects have roughly the correct size, abundance, star
formation rate and internal characteristic velocity, but it is not yet
clear whether they are indeed disk-like. Very recent studies of the
abundance of such objects during $0<z<5$ suggest that
the overall rate of unobscured star formation in the Universe actually peaked
at $1<z<2$, that star formation in this mode could possibly account
for {\it all} the observed stars in galaxies, and that most 
stars formed after $z\sim 1$ (Madau et al 1997).
Such late star formation is one of the most robust and controversial 
predictions of hierarchical models (e.g. White 1989; Cole et al 1994)
but observational verification is difficult
since the conversion from observed UV flux to star-formation rate is
uncertain by at least a factor of 2. Thus one cannot tell
whether all stars formed in the observed unobscured mode
or only 30\% to 50\% of them, for example the stars in
present-day disks. A direct proof of recent elliptical
formation could come from a survey of the Universe at, 
say, $z=2$, which showed the current population to be absent at that epoch.
Recent deep redshift surveys selected at I and K
allow complete samples of early-type galaxies to be
identified to $z\sim 1$. 
$V/V_{max}$ tests applied to these samples
show unambiguously that the early-type population does not
follow standard passive evolution models (Kauffmann et
al 1996). In fact, roughly two thirds of the present population
appears to be missing at $z=1$; either the galaxies were
actively forming stars or they were in several pieces at that time.
If further deep surveys confirm this result,
we may conclude that the bulk of galaxy formation has
already been observed, and that we now have a crude quantitative
understanding of the origin and evolution of the basic properties of
galaxies. 
\medskip

\noindent{{\bf Acknowledgements} I thank my collaborators S. Charlot
G. Kauffmann, S. Mao and H.J. Mo for many helpful discussions of the
material in this review. G. Kauffmann also provided the model data
plotted in figures 1 and 2.

%
%

\end{document}